# Resolving Polarization Switching Pathways of Sliding Ferroelectricity in Trilayer 3R-MoS$_2$

Jing Liang[1,2,#], Dongyang Yang[1,2,#], Jingda Wu[1,2], Yunhuan Xiao[1,2], Kenji Watanabe[3], Takashi Taniguchi[4], Jerry I Dadap[1,2], Ziliang Ye[1,2]*

[1] Quantum Matter Institute, The University of British Columbia, Vancouver, BC V6T 1Z4, Canada

[2] Department of Physics and Astronomy, The University of British Columbia, Vancouver, BC V6T 1Z1, Canada

[3] Research Center for Functional Materials, National Institute for Materials Science, 1-1 Namiki, Tsukuba 305-0044, Japan

[4] International Center for Materials Nanoarchitectonics, National Institute for Materials Science, 1-1 Namiki, Tsukuba 305-0044, Japan

\# These authors contributed equally to this manuscript.

* Correspondence: zlye@phas.ubc.ca

**Abstract**

Exploring the pathways of polarization switching in 2D sliding ferroelectrics with multiple internal interfaces is crucial for understanding the switching mechanism and for enhancing their performance in memory-related applications. However, distinguishing the rich configurations of various stacking from a coexistence of polarization domains has remained challenging. In this investigation, we employ optical techniques to resolve the stacking degeneracy in a trilayer 3R-MoS$_2$ across several polarization switching cycles. Through a comprehensive analysis of the unique excitonic response exhibited by different layers, we unveil multiple polarization switching pathways that are determined by the sequential release of domain walls initially pinned at various interfaces within the trilayer, providing an understanding of the switching mechanism in multilayered sliding ferroelectrics. Our study not only reveals the intricate dynamics of polarization switching, but also underscores the crucial role of controlling domain walls, pinning centers, and doping levels, offering new insights for enhancing the applications of these materials in sensing and computing.

**Main**

The stacking order in 2D materials has introduced a new dimension for exploring novel phenomena and functional devices. Notably, sliding ferroelectricity has recently been observed in rhombohedral-stacked transition metal dichalcogenides (R-TMDs), both artificially stacked and chemically synthesized, where a spontaneous electrical polarization arises from an asymmetric interlayer-coupling-induced Berry phase[1-13]. Unlike conventional ferroelectric materials, polarization switching in sliding ferroelectricity involves the collective motion of atomic layers along the in-plane direction, perpendicular to the driving electric field[14-17].

When the stacking is coherent among layers, the polarization-induced interlayer potential can accumulate in multilayer 3R-TMDs, which are semiconductors with sub-eV bandgaps and strong excitonic effects, leading to novel applications in sensing and computing[18-20]. Conversely, incoherent stacking allows an $n$-layer material to host $2^{n-1}$ polarization configurations, significantly expanding the phase space for applications related to information storage[6-8]. However, electrically distinguishing various intermediate stacking configurations from a coexistence of polarization domains has proven challenging. Here we show that the configuration of an incoherently stacked 3R-TMD trilayer can be resolved by leveraging the unique excitonic response of each substituent layer. In particular, we use reflection contrast spectroscopy with a diffraction-limited resolution to fully identify the stacking configuration of a trilayer 3R-MoS$_2$ in its initial, intermediate, and final states during multiple polarization switching cycles.

The polarization switching process in bilayer 3R-TMDs has been extensively studied. When subjected to an out-of-plane electric field, the domain of the polarization along the same direction experiences a lower free energy and tends to expand, while the domain of the opposite polarization tends to contract, thus creating a driving force for the movement of the domain wall (DW) [4,5,9-11]. Upon surpassing the free energy difference relative to the trapping potential of the pinning center that initially localizes the DW, the DW becomes depinned, thus triggering the polarization switching. The switch can be non-volatile if the DW finally becomes pinned again[10].

A similar process is expected in trilayer 3R-TMDs, introducing two interfacial polarizations and four potential polarization configurations. Specifically, we designate states with aligned interfacial polarizations as ABC and CBA stacking, and those anti-aligned interfacial polarizations as ABA and BAB stacking. Theoretical predictions suggest that one of the anti-aligned states (ABA) may be energetically favored during the transition from ABC to CBA stacking, with the other anti-aligned state (BAB) energetically favourable in the reverse process. However, such a contrast has not been discerned by the transport measurements due to the degeneracy in the net polarization[6].

In contrast to electrical measurement, optical spectroscopy offers the capability to resolve the stacking configuration of different polarization states based on their distinctive excitonic responses. In bilayer 3R-$MoS_2$, the broken symmetry between the top and bottom layer, results in different band edges, equivalent to a type-II alignment at the K point[3,21]. Upon optical excitation, photocarriers migrate to the layer with a lower potential, forming an interlayer exciton whose dipole orientation can indicate the stacking order. Moreover, the intralayer excitons in the two layers exhibit approximately a 10-meV difference in peak energy, and a comparison of their oscillator strengths with finite doping can further reveal the stacking configuration[11]. Since the interlayer exciton yields identical responses between ABA and BAB stackings, here we employ reflection contrast (RC) spectroscopy to detect polarization switching events and identify stacking configurations in a trilayer 3R-$MoS_2$. In analogy to the bilayer, we expect a trilayer with aligned polarizations will exhibit an RC spectrum having three excitonic peaks, reflecting the diverse chemical environments of the three layers[3,22-26]. Conversely, if the polarizations become anti-aligned, the two outer layers become symmetric, resulting in a spectrum similar to that of the bilayer.

We investigate the RC spectra of a 3R-$MoS_2$ trilayer with double-side encapsulation and with both top and bottom graphene gates that provide control over doping level and electric field (Figure 1a; Methods in Supplementary Information). All measurements are performed at 1.6 K. In Figure 1b, the field dependence of the first energy derivative of the RC spectra (dRC) is presented as the electric field is scanned from the positive to negative limit. Three intralayer excitonic peaks at approximately 1.9 eV are observed at the positive limit, suggesting an initial polarization state to either ABC or CBA stacking, consistent with the 3R bilayer analogy.

At $E_1 = -0.064\ V/nm$, a sharp transition in the dRC spectra occurs, with an increased separation between excitonic peaks and a reduction in peak number to two, indicating a change in stacking configuration to either ABA or BAB. Subsequently, at $E_2 = -0.087\ V/nm$, a second pronounced change in the exciton energy signifies another switch to a third state. Similar to the initial state, a three-peak dRC spectrum is observed in the final state. Besides the backward scan, we perform a forward electric field scan, where we observe two analogous transitions at two positive fields ($E_3 = 0.084\ V/nm,\ E_4 = 0.100\ V/nm$). Importantly, the immediate state exhibits similar spectroscopic features as in the backward scan: two excitonic peaks with the comparable peak splitting and oscillator strength ratio. Consequently, we conclude that the polarization switching pathways in the forward and back directions involve the same intermediate state in this device, contrary to the previous prediction of one intermediate state being more favorable than the other.

The hysteresis loop depicting the two-stage switching process observed in device 1 is illustrated in Figure 1c. In the positive and negative field limits, the stacking configurations are identified as CBA and ABC, respectively. The common intermediate state in this process is an ABA state, where the two anti-aligned interfacial polarizations give rise to a zero net polarization. The single hysteresis loop suggests that such an intermediate state is metastable, indicating the absence of anti-ferroelectric phase in our system. However, the absence of BAB states in this scan cycle does not preclude their possibility. Figure 4 demonstrates opposite intermediate states in different cycles of electric field training, and we attribute this diversity in the switching pathway to the competition in trapping potential among pinning centers in different layers.

The stacking configuration of different polarization states can be identified by analyzing their optical response as a function of electric field and doping. In the backward scan of Figure 1b, the middle peak diminishes as the electric field is swept from positive to zero, and the low energy peak intersects with a rapidly redshifting peak with increasing negative field, just before the stacking switches at (1). These spectral features agree well with our observations in an unswitchable CBA-stacked trilayer (Figure 2a). Similar to the bilayer device, not every trilayer device exhibits switchability, which we attribute to the absence of pre-existing DWs.

As depicted in Figure 2a, two avoided-crossing features emerge in the lower-energy peak, occurring at $-0.05\ V/nm$ and $-0.12\ V/nm$. These features are identified as resonances between the intralayer exciton and the interlayer exciton ($IX^-$) transition, involving the valence band in one layer and conduction band in the other (Figure 2b). When the external electric field is anti-aligned with the spontaneous polarization direction, the interlayer potential increases and the band offsets between layers become larger, thus causing a redshift in the interlayer exciton. The out-of-plane electric dipole moments, determined from the Stark shift slopes are approximately 0.62 $e$·nm. Interlayer excitons, characterized by a smaller binding energy and weaker oscillator strength compared to intralayer excitons, cannot be observed in the RC spectra until they hybridize with the intralayer species. Since the avoided crossings are exclusively observed in the negative field range, we conclude the initial stacking order is coherent among the three layers, defined as the CBA stacking. Between the two interlayer excitons, a peak splitting of approximately 46 meV is observed, similar to that observed in 3R-$MoS_2$ bilayers (Figure S1). This splitting significantly exceeds the spin-orbit-induced spin splitting in the conduction band. Furthermore, the low-energy interlayer exciton can only hybridize with the low-energy intralayer exciton. These features suggest that phonon-assisted intervalley interlayer transitions may play a crucial role, warranting further investigations[26-30].

Similar to the initial state, the stacking configuration of the final state in Figure 1b is identified as ABC by comparing the spectra with another unswitchable device (Figure 2d). Both devices show avoid-crossing features between the intralayer exciton and interlayer exciton at approximately $0.05\ V/nm$. The interlayer exciton redshifts with a positive field, indicating the band alignment is opposite to the initial state. The missing inter-intra exciton crossing in Figure 1b also confirms the switching in the stacking configuration. In contrast to CBA, the ABC stacking displays three peaks at the negative field limit (Figure 2c, f) and the middle peak gradually diminishes as the external field is swept positively.

To explore the layer origin of three intralayer excitons and to confirm the stacking assignment, we analyze the electric field dependence of a CBA-stacked device at a fixed electron doping density of $1.4 \times 10^{12}\ cm^{-2}$ (Figure 2g). When the external field is small ($E < 0.077\ V/nm$), the electrons reside in the bottom layer, resulting in the observation of excitons in the top and middle layers ($X_t$, $X_m$), and the attractive and repulsive exciton polarons ($XP_b^{-\prime}$, $XP_b^-$) in the bottom layer (left panel in Figure 2h), similar to the RC spectrum with finite doping and without electric field (Figure S2).

As the electric field increases, the conduction band offset $\Delta_c^t$ (between the top and middle layers) and $\Delta_c^b$ (between the middle and bottom layers) decrease until the Fermi level reaches the conduction band edge of the middle layer ($E = 0.077\ V/nm$). From this point on, electrons begin to migrate from the bottom layer to the middle layer (middle panel in Figure 2h), which leads to the formation of attractive and repulsive exciton polarons, $XP_m^{-\prime}$ and $XP_m^-$, in the middle layer. Due to the similar energy of $XP_m^-$ and $X_t$, $XP_m^-$ only enhances the oscillator strength of $X_t$ (Figure S2). Simultaneously, $XP_b^-$ experiences a redshift and transfers its oscillator strength to the intrinsic $X_b$. As the electric field continues to increase, $\Delta_c^t$ decreases further while $\Delta_c^b$ is reversed. At $E = 0.152\ V/nm$, the Fermi level reaches the conduction band edge of the top layer, causing electrons migration from the middle layer to the top layer (right panel in Figure 2h). This change causes $XP_m^{-\prime}$ to blueshift and transfer its oscillator strength to intrinsic $X_m$. The interaction between $X_t$ and the Fermi surface also results in the formation of attractive and repulsive exciton polaron in the top layer, enhancing the oscillator strength in $X_m$ and $X_b$ due to their comparable energies (Figure S2).

Overall, the field dependent spectra support our stacking assignment and suggests the three exciton peaks, from low to high energy, originate from the middle, top, and bottom layer respectively in the CBA stacking. Interestingly, the two-staged transition suggests that the intrinsic conduction band offset $\Delta_c^t$ is larger than $\Delta_c^b$. Additionally, the oscillator strength of the highest-energy $X_b$ peak decreases as the positive field increases (Figure 2a). This change can be attributed to field-assisted exciton dissociation: as the band offset $\Delta_c^b$ decreases with increasing positive field, it facilitates exciton dissociation through electron transfer from the bottom layer to the middle layer, leading to a decrease in the oscillator strength of $X_b$. Opposite changes can be observed in the middle $X_t$ peak[31,32].

In contrast to the initial and final state, the immediate state exhibits remarkable differences in the RC spectra, where two excitonic peaks are observed with a larger separation of approximately 34 meV. Figure 3 is based on a stable intermediate domain, where switching to either ABC or CBA requires a large coercive field, which allows us to study the doping and field dependence within the presented range. Two excitonic peaks are observed at 1.887 eV and 1.921 eV, respectively. Field dependence measurements in this device reveal two pairs of avoided crossing features, with one pair occurring at positive fields and the other at negative fields. These avoided crossing features resemble the spectra in both CBA and ABC stacked devices, suggesting they originate from two types of interlayer excitons with opposite dipole moments. This corresponds to the band alignment in an incoherent stacking of either ABA or BAB.

In either ABA or BAB stacking configurations, there exist two types of intralayer excitons, which we label as $X_A$ from the A layer and $X_B$ from the B layer. Previous 3R-bilayer study has established that $X_A$ exhibits higher energy than $X_B$[3]. In trilayer configurations, the A and B layers are in different dielectric environments (Figure S3). In ABA stacking, the B layer is sandwiched by two MoS$_2$ layers, while the A layer is adjacent to one MoS$_2$ layer and one BN layer. Since the dielectric constant of MoS$_2$ is larger than BN, $X_B$ experiences more screening than $X_A$, resulting in a relative redshift of $X_B$ and an increase in the exciton peak splitting. Conversely, in the BAB configuration, the A layer undergoes stronger screening, causing a redshift in the high-energy exciton peak and reducing the peak splitting. If this peak splitting is smaller than the linewidth, the spectral response effectively merges into a single peak that comprises all oscillator strengths. Therefore, we attribute the observed intermediate state in Figure 1, with significant splitting, to ABA, and the intermediate state in Figure 4d, characterized by a single strong excitonic peak, to BAB. The large oscillator strength of the single excitonic peak in the BAB configuration and the ratio of the oscillator strength between the two peaks in the ABA configuration all corroborate our assignment.

The band alignment at the K point among three layers in ABA stacking is illustrated in Figure 3b, where conduction band in the middle layer is higher than the two outer layers. Under an external electric field, the conduction band offset between the middle layer and one of the outer layers decreases, leading to a redshift in one pair of interlayer excitons, as observed in Figure 3b and 3c. This band alignment is further confirmed by the doping dependence in Figure 3d, where the responses can be classified into three regions, electron-doped (I), intrinsic (II), and hole-doped (III) regions.

In region-II, where the Fermi level lies inside the bandgap of all layers, the spectrum exhibits two intrinsic excitonic peaks. In region-I, the trilayer is doped by electrons, and the doped electrons should reside in the top and bottom layers according to the band alignment (Figure 3e). Experimentally, the high-energy peak transforms into a blueshifted branch and a redshifted branch, known as repulsive and attractive exciton polaron ($XP_{t,b}^-$ and $XP_{t,b}^{-\prime}$), respectively. The low-energy peak remains largely unchanged, except for some redshift at large doping, likely due to the screening effect. In region-III, the trilayer is hole-doped at the Γ point (Figure 3f, S4), where the three layers strongly hybridize, converting all excitons into exciton-polaron peaks with varying redshifts in different layers. The binding energy differences (16 meV of $XP_m^+$, and 4 meV of $XP_t^{+\prime}$ and $XP_b^{+\prime}$, respectively) may arise from finite layer polarization-induced unequal hole distribution between three layers at the Γ point.

To investigate the polarization switching mechanism, we conduct the electric field on the same device as in Figure 1 for multiple cycles, where we observe the switching pathway and coercive fields can vary. For example, in the backward scan of a different cycle (Figure 4a), the spectrum is similar as in Figure 3a, indicating the same ABA stacking in the intermediate state, except that the switching fields become different. On the other hand, in the intermediate state of the forward scan (Figure 4c), the RC spectrum exhibits a large peak at 1.918 eV, which we attribute to the intermediate state with BAB stacking. The hysteresis loop illustrating the two-stage switch in Figure 4 is summarized in 4e. In the positive and negative field limits, the stacking configuration is CBA and ABC, respectively, while the intermediate state is BAB in the forward scan and ABA in the backward scan. The two-stage switch occurs as the external electric field provides a driving force to depin the domain wall (DW) at different interface sequentially, as depicted in Figure 4b.

Initially, the sample has two pre-existing DWs located at the top and bottom interfaces. These DWs are trapped by pinning centers such as defects and bubbles outside the optical probing area, resulting in a homogenous CBA response. When the external field reaches $E_1 = -0.043 \ V/nm$, the top domain wall (DW$_t$) is released from the pinning center with a weaker pinning potential $p_t$, causing a local switch from CBA to ABA. Subsequently, at $E_2 = -0.087 \ V/nm$, the bottom domain wall (DW$_b$) is released from the pinning centers with a stronger pinning potential $p_b$, causing ABA to switch to ABC. The difference between $p_t$ and $p_b$ determines the electric field window where the intermediate state ABA can be observed. The sequential release of DW$_t$ and DW$_b$, which form the intermediate ABA state, is supported by the electric-field-dependent reflection contrast map that indicates the distribution of domains and domain walls (Figure S5, 6).

After being released from the initial pinning centers on one side of the optical probe, the DW can quickly sweep through the focus spot, and finally becomes trapped again by the pinning centers on the other side of the probe. Consequently, the sequence of the DW release and the intermediate state's stacking in the reverse scan depend on the strength of the final pinning centers ($p_t'$ and $p_b'$). In the scan cycle in Figure 4, $p_t'$ is smaller than $p_b'$, and therefore DW$_t$ is released earlier than DW$_b$, causing ABC to switch back to CBA through a BAB intermediate state. However, in the scan cycle in Figure 1, $p_t'$ is larger than $p_b'$, and the intermediate state becomes ABA (Figure S7).

Experimentally, our devices exhibit additional switching pathways, and we can attribute these processes to different combinations of pinning potentials ($p_t$ and $p_b$). In Figure S8, we illustrate a single sharp transition between CBA and ABC stacking without an observable intermediate state. This behavior is consistent with the scenario where the pinning potentials in the upper and lower interfaces are comparable, i.e., $p_t \sim p_b$ and $p_t' \sim p_b'$. In another trilayer device, we observed different spots where the switch occurs exclusively between CBA and ABA (Figure S9) or only between ABA and ABC (Figure S10). These observations suggest that some pinning centers in one layer can be exceptionally strong, thereby allowing only the DW in the other interface to sweep through the probe spot.

To understand the variation in the polarization switching pathway, we performed statistical analysis on the coercive fields, which are summarized in Figures S11 and S12. Clearly, the coercive field varies up to 30% between cycles and different devices can exhibit very different switching pathway statistics. We attribute such a stochastic behavior to the presence of a network of pinning centers with different potentials – despite having the same electric field loop applied in each cycle, domain walls are not deterministically pinned in the same configuration[10,33-37].

In addition, statistical analysis shows that the ABA stacking appears much more frequently than the BAB stacking. (BAB stacking is only observed occasionally in the forward scan of one device, out of five switchable devices.) Rather than to the energy difference between two intermediate stackings, here we attribute the prevalence of ABA stacking to an asymmetric screening due to an n-type initial doping. When a thin layer of semiconductor is placed in a uniform electric field, the electronic band structure near both surfaces will be bent within the space charge layer. Similar to the Debye length, the space charge layer thickness is determined by the local density of states (DOS) at the Fermi level. If the semiconductor is chemically n-type doped with the Fermi level close to the conduction band edge, an electric field out of the film that induces hole doping can cause a much larger band bending than an electric field into the film that causes an additional electron doping, because the DOS increases significantly if the Fermi level enters the conduction band. As a result, one surface can experience much more band bending and admit more electric field than the other surface.

Experimentally, we observed a large contrast in the screening of two gate fields (Figure 5): when the trilayer is in CBA stacking, only the top gate that generates an electric field out of the film can achieve the switching; the bottom gate appears mostly screened by the A layer. The Fermi level in this device is clearly close to the conduction band edge, as reflected in the early appearance of the polaron peak (Figure 5b), and such an n-type doping is common among $MoS_2$ crystals. Mostly comprising the electric field from top gate, the total displacement field in the CBA stacked MoS$_2$ is therefore stronger at the upper interface than at the lower interface (Figure 5a). If the pinning centers at both interfaces have similar trapping potentials, the domain wall at the upper interface will then experience a larger stress and become unpinned first. As a result, the CBA stacking will switch to ABA first and then switch to ABC.

In the opposite switching direction from ABC to CBA, both the external field direction and screening asymmetry are reversed (Figure 5c, d). The total displacement field thus becomes stronger at the lower interface than at the upper interface and moves the bottom domain wall first. Consequently, ABA is always the favored intermediate state whichever the switching direction is. Since the external field cannot induce layer-polarized carriers in either intrinsic or p-type doped flakes for efficient screening, we expect the displacement fields at the two interfaces to be similar in these scenarios, and neither the ABA nor the BAB intermediate state will be favored over the other. Overall, such an asymmetric screening effect should play a significant role in any multilayered devices with n-type doping and the switching pathway in a 3R-MoS$_2$ device is jointly determined by many extrinsic factors including the pinning center and domain wall distribution, as well as the doping that can screen the external field.

In conclusion, our investigation into the excitonic response of 3R-MoS$_2$ trilayers under varying electric fields and doping carrier densities has allowed us to distinguish different stacking configurations in a trilayer, even when net polarizations are identical. This resolution enables the identification of possible pathways during polarization switches, revealing that intermediate states are primarily determined by the sequence of DW releases and, therefore, the relative strength of pinning centers at distinct interfaces that initially localize the DWs. Variability in switching behavior across multiple scan cycles indicates a potential impact of DW motion on the distribution of pinning centers. Moreover, we found that one intermediate state occurs more frequently than the other, which can be attributed to an asymmetric screening effect arising from the n-type initial doping. Although this study is focused on the 3R-MoS$_2$ trilayer as a prototype, our insights should be applicable to other sliding ferroelectric material systems with multiple interfaces[38].

**Acknowledgement**

We acknowledge support from the Natural Sciences and Engineering Research Council of Canada, Canada Foundation for Innovation, New Frontiers in Research Fund, Canada First Research Excellence Fund, and Max Planck–UBC–UTokyo Centre for Quantum Materials. Z.Y. is also supported by the Canada Research Chairs Program. K.W. and T.T. acknowledge support from JSPS KAKENHI (Grant Numbers 19H05790, 20H00354 and 21H05233).

**Author Contributions**

J.L. and Z.Y. conceived the work. J.L. and D.Y. fabricated the samples. J.L. and D.Y. conducted the measurements with the help from J.W., Y.X. and J.I.D.. J.L. and Z.Y. analyzed the data. K.W. and T.T. provided the hBN crystal. Z.Y. supervised the project. J.L., Z.Y., and J.I.D. wrote the manuscript based on the input from all other authors. J.L. and D.Y. contributed equally to this work.

**Competing interests**

The authors declare no competing interests.
13 / 21

Figures and captions:

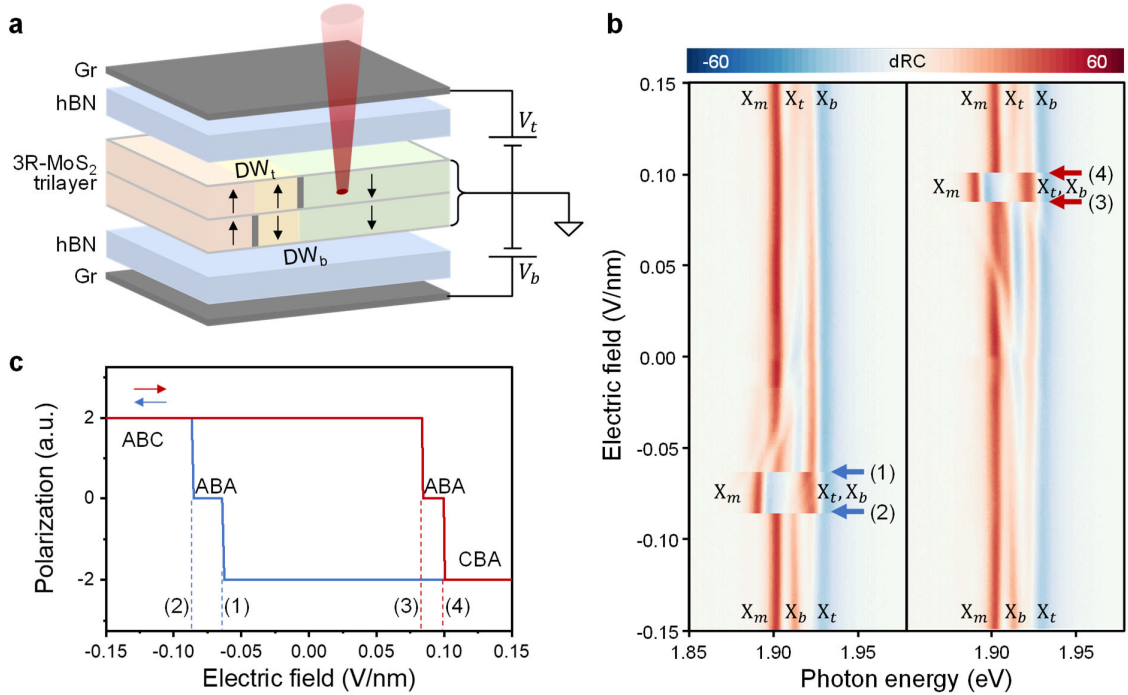

**Figure 1 | Polarization switching in a 3R-MoS$_2$ trilayer involving ABA stacking as the intermediate state.** (**a**) Schematic of reflectance contrast (RC) spectrum measurement in a dual-gated trilayer device. The arrows indicate the polarization direction at each interface. Three domains of distinct net polarization are separated by one domain wall located at the top interface (DW$_t$) and the other at the bottom interface (DW$_b$). (**b**) Electric field-dependent first energy derivative of the RC (dRC) spectra in the backward (left panel, from positive to negative field) and forward scan (right panel, from negative to positive field). The blue and red arrows highlight the polarization switching events. (**c**) The net polarization as a function of electric field in the forward (red) and backward (blue) scan directions, showing a hysteretic behavior.

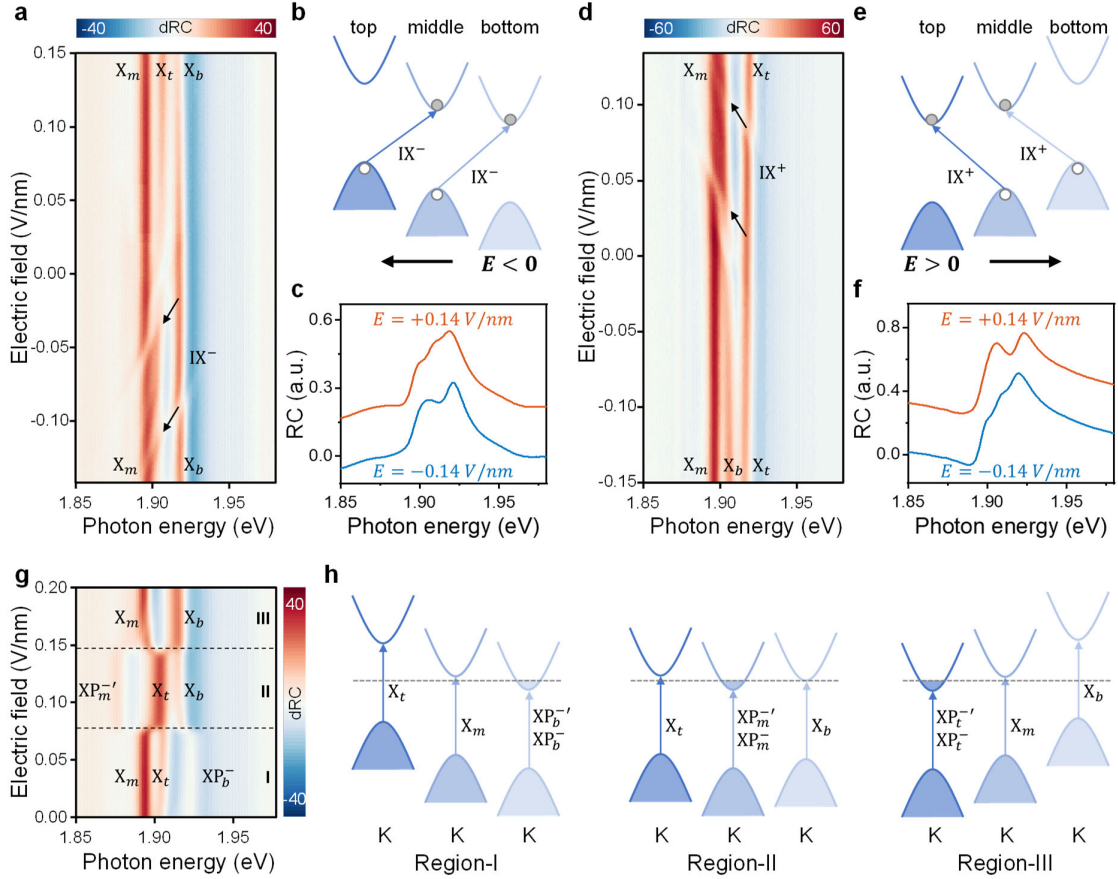

**Figure 2 | Determining the stacking configuration of the initial and final states during the polarization switch.** (**a**) Electric field-dependent dRC spectra of a sample with CBA stacking. The arrows indicate interlayer excitons ($IX^-$) emerging at negative field. $X_i$ represents the intrinsic intralayer exciton. The subscript *i* represents *t*, *m*, or *b*, denoting the optical transitions in the top, middle, or bottom layers. (**b**) Schematics of interlayer transitions in CBA stacking under a negative electric field. (**c**) RC spectrum of CBA stacking at largest positive (red) and negative fields (blue), which are similar as in Figure 1(b). (**d-f**) are similar as **a-c** but from a sample with ABC stacking, whose polarization is opposite to CBA. (**g**) Electric field-dependent dRC spectra of CBA stacking at a fixed electron doping density. The dashed lines divide the spectrum into three regions where the electrons are doped into: bottom layer (region-I), middle layer (region-II), and top layer (region-III). (**h**) Schematics of the band alignments and optical transitions in CBA stacking with a fixed electron doping residing in different layers. The gray dashed lines denote the Fermi level position.

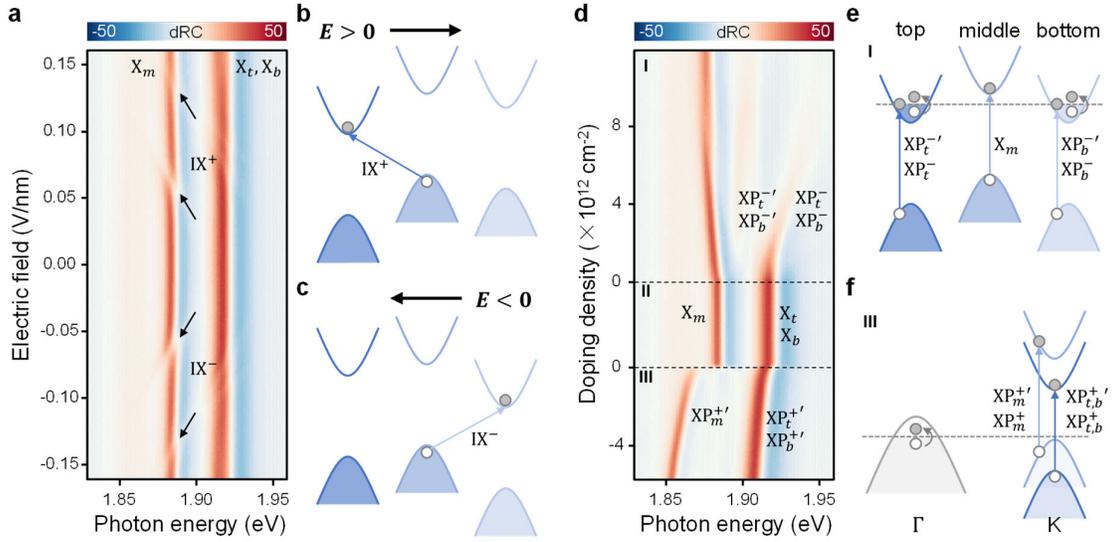

**Figure 3 | Determining the stacking configuration of the intermediate state. (a)** Electric field-dependent dRC spectra of a sample with ABA stacking. Unlike ABC or CBA stacking, two pairs of interlayer excitons are observed and labeled as $IX^+$ and $IX^-$. **(b, c)** Band alignments at K points and interlayer transitions in ABA stacking under a positive (b) and negative (c) electric field. **(d)** Doping-dependent dRC spectrum of the same sample. Two lines divide the spectrum into three regions, which is intrinsic or electron- or hole-doped. **(e, f)** Band alignments corresponding to region I and III in **(d)**. The gray dashed lines denote the Fermi level.

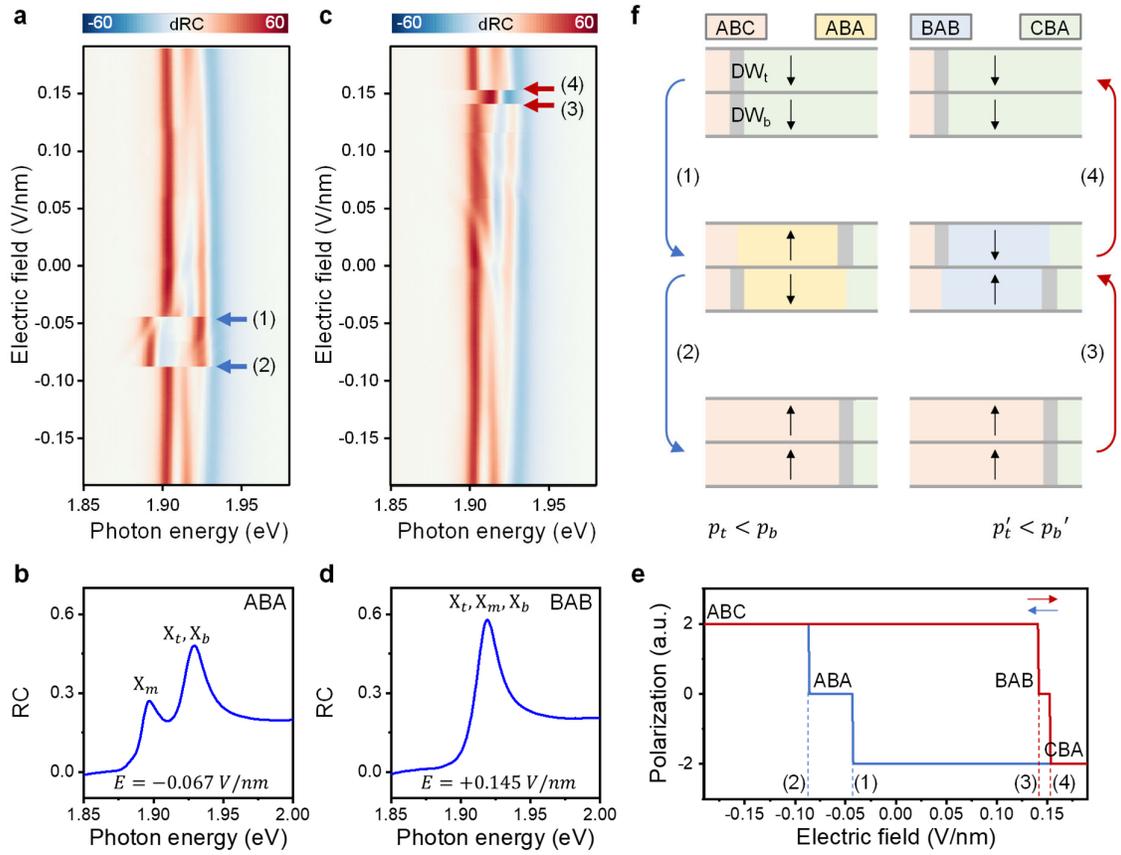

**Figure 4 | Resolving polarization switching pathways with ABA and BAB stacking as intermediate states. (a & c)** Electric field-dependent dRC spectrum of the Figure 1 sample in a different scan cycle. The intermediate states, which are between the coercive field (1) and (2), and between (3) and (4), clearly exhibit distinctive optical responses. **(b & d)** RC spectra of the two intermediate states. **(e)** The net polarization as a function of electric field in the forward (red) and backward (blue) scan directions, showing a different hysteretic behavior compared to Figure 1. **(f)** Schematics of the polarization switching pathways in this cycle. The green, orange, yellow, and blue regions correspond to CBA, ABC, ABA, and BAB stacking, respectively.

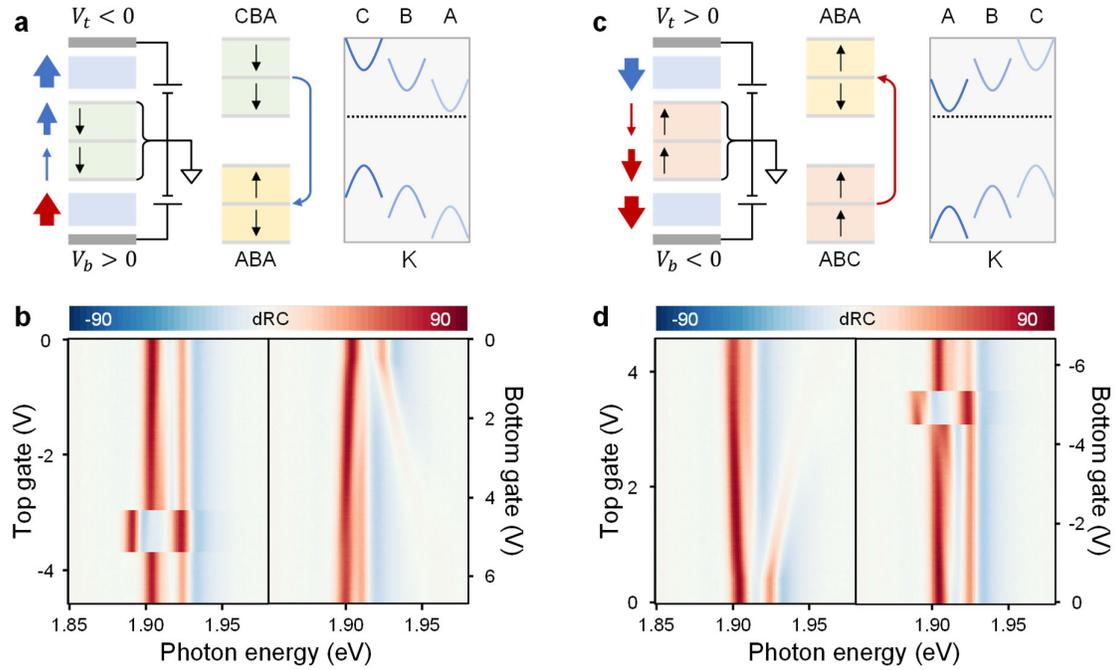

**Figure 5 | Single-gate-dependent polarization switching pathways.** (a & c) Illustration of the external electric field contributions from the top (blue arrows) and bottom (red arrows) gates to the 3R-MoS$_2$ trilayer, with the initial state being CBA (a) and ABC (c) polarization states, respectively. The middle panel depicts the first switching pathway, while the right panel illustrates the band structure of the initial polarization state. (b & d) Opposite switching behaviors due to the asymmetric screening effect for initial states of CBA (b) and ABC (d) stackings, respectively.

# References

(1) Li, L. & Wu, M. Binary Compound Bilayer and Multilayer with Vertical Polarizations: Two-Dimensional Ferroelectrics, Multiferroics, and Nanogenerators. *ACS Nano* **2017**, *11*, 6382-6388.

(2) Wu, M. & Li, J. Sliding ferroelectricity in 2D van der Waals materials: Related physics and future opportunities. *Proceedings of the National Academy of Sciences* **2021**, *118*, e2115703118.

(3) Liang, J., Yang, D., Wu, J., Dadap, J. I., Watanabe, K., Taniguchi, T. & Ye, Z. Optically Probing the Asymmetric Interlayer Coupling in Rhombohedral-Stacked $MoS_2$ Bilayer. *Physical Review X* **2022**, *12*, 041005.

(4) Weston, A., Castanon, E. G., Enaldiev, V., Ferreira, F., Bhattacharjee, S., Xu, S., Corte-León, H., Wu, Z., Clark, N., Summerfield, A. *et al.* Interfacial ferroelectricity in marginally twisted 2D semiconductors. *Nature Nanotechnology* **2022**, *17*, 390-395.

(5) Wang, X., Yasuda, K., Zhang, Y., Liu, S., Watanabe, K., Taniguchi, T., Hone, J., Fu, L. & Jarillo-Herrero, P. Interfacial ferroelectricity in rhombohedral-stacked bilayer transition metal dichalcogenides. *Nature Nanotechnology* **2022**, *17*, 367-371.

(6) Meng, P., Wu, Y., Bian, R., Pan, E., Dong, B., Zhao, X., Chen, J., Wu, L., Sun, Y., Fu, Q. *et al.* Sliding induced multiple polarization states in two-dimensional ferroelectrics. *Nature Communications* **2022**, *13*, 7696.

(7) Deb, S., Cao, W., Raab, N., Watanabe, K., Taniguchi, T., Goldstein, M., Kronik, L., Urbakh, M., Hod, O. & Ben Shalom, M. Cumulative polarization in conductive interfacial ferroelectrics. *Nature* **2022**, *612*, 465-469.

(8) Liang, J., Yang, D., Xiao, Y., Chen, S., Dadap, J. I., Rottler, J. & Ye, Z. Shear Strain-Induced Two-Dimensional Slip Avalanches in Rhombohedral $MoS_2$. *Nano Letters* **2023**, *23*, 7228-7235.

(9) Yang, D., Liang, J., Wu, J., Xiao, Y., Dadap, J. I., Watanabe, K., Taniguchi, T. & Ye, Z. Non-volatile electrical polarization switching via domain wall release in 3R-$MoS_2$ bilayer. *Nature Communications* **2024**, *14*, 1389.

(10) Ko, K., Yuk, A., Engelke, R., Carr, S., Kim, J., Park, D., Heo, H., Kim, H.-M., Kim, S.-G., Kim, H. *et al.* Operando electron microscopy investigation of polar domain dynamics in twisted van der Waals homobilayers. *Nature Materials* **2023**, *22*, 992-998.

(11) Molino, L., Aggarwal, L., Enaldiev, V., Plumadore, R., I. Fal´ko, V. & Luican-Mayer, A. Ferroelectric Switching at Symmetry-Broken Interfaces by Local Control of Dislocations Networks. *Advanced Materials* **2023**, *35*, 2207816.

(12) Ji, J., Yu, G., Xu, C. & Xiang, H. J. General Theory for Bilayer Stacking Ferroelectricity. *Physical Review Letters* **2023**, *130*, 146801.

(13) Wang, L., Qi, J., Wei, W., Wu, M., Zhang, Z., Li, X., Sun, H., Guo, Q., Cao, M., Wang, Q.

*et al.* Bevel-edge epitaxy of ferroelectric rhombohedral boron nitride single crystal. *Nature* **2024**, *629*, 74-79.

(14) Bennett, D. Theory of polar domains in moiré heterostructures. *Physical Review B* **2022**, *105*, 235445.

(15) Bennett, D. & Remez, B. On electrically tunable stacking domains and ferroelectricity in moiré superlattices. *npj 2D Materials and Applications* **2022**, *6*, 7.

(16) Enaldiev, V. V., Ferreira, F. & Fal'ko, V. I. A Scalable Network Model for Electrically Tunable Ferroelectric Domain Structure in Twistronic Bilayers of Two-Dimensional Semiconductors. *Nano Letters* **2022**, *22*, 1534-1540.

(17) Bennett, D., Chaudhary, G., Slager, R.-J., Bousquet, E. & Ghosez, P. Polar meron-antimeron networks in strained and twisted bilayers. *Nature Communications* **2023**, *14*, 1629.

(18) Yang, D., Wu, J., Zhou, B. T., Liang, J., Ideue, T., Siu, T., Awan, K. M., Watanabe, K., Taniguchi, T., Iwasa, Y. *et al.* Spontaneous-polarization-induced photovoltaic effect in rhombohedrally stacked $MoS_2$. *Nature Photonics* **2022**, *16*, 469-474.

(19) Wu, J., Yang, D., Liang, J., Werner, M., Ostroumov, E., Xiao, Y., Watanabe, K., Taniguchi, T., Dadap, J. I., Jones, D. *et al.* Ultrafast response of spontaneous photovoltaic effect in 3R-$MoS_2$-based heterostructures. *Science Advances* **2022**, *8*, eade3759.

(20) Dong, Y., Yang, M.-M., Yoshii, M., Matsuoka, S., Kitamura, S., Hasegawa, T., Ogawa, N., Morimoto, T., Ideue, T. & Iwasa, Y. Giant bulk piezophotovoltaic effect in 3R-$MoS_2$. *Nature Nanotechnology* **2023**, *18*, 36-41.

(21) Sung, J., Zhou, Y., Scuri, G., Zólyomi, V., Andersen, T. I., Yoo, H., Wild, D. S., Joe, A. Y., Gelly, R. J., Heo, H. *et al.* Broken mirror symmetry in excitonic response of reconstructed domains in twisted $MoSe_2$/$MoSe_2$ bilayers. *Nature Nanotechnology* **2020**, *15*, 750-754.

(22) Mak, K. F., Lee, C., Hone, J., Shan, J. & Heinz, T. F. Atomically Thin $MoS_2$: A New Direct-Gap Semiconductor. *Physical Review Letters* **2010**, *105*, 136805.

(23) Mak, K. F., He, K., Lee, C., Lee, G. H., Hone, J., Heinz, T. F. & Shan, J. Tightly bound trions in monolayer $MoS_2$. *Nature Materials* **2013**, *12*, 207-211.

(24) Raja, A., Chaves, A., Yu, J., Arefe, G., Hill, H. M., Rigosi, A. F., Berkelbach, T. C., Nagler, P., Schuller, C., Korn, T. *et al.* Coulomb engineering of the bandgap and excitons in two-dimensional materials. *Nat Commun* **2017**, *8*, 15251.

(25) Wang, G., Chernikov, A., Glazov, M. M., Heinz, T. F., Marie, X., Amand, T. & Urbaszek, B. Colloquium: Excitons in atomically thin transition metal dichalcogenides. *Reviews of Modern Physics* **2018**, *90*, 021001.

(26) Leisgang, N., Shree, S., Paradisanos, I., Sponfeldner, L., Robert, C., Lagarde, D., Balocchi, A., Watanabe, K., Taniguchi, T., Marie, X. *et al.* Giant Stark splitting of an exciton in bilayer $MoS_2$. *Nature Nanotechnology* **2020**, *15*, 901-907.

(27) Yu, H., Liu, G.-B. & Yao, W. Brightened spin-triplet interlayer excitons and optical

selection rules in van der Waals heterobilayers. *2d Mater* **2018**, *5*, 035021.

(28) Robert, C., Han, B., Kapuscinski, P., Delhomme, A., Faugeras, C., Amand, T., Molas, M. R., Bartos, M., Watanabe, K., Taniguchi, T. *et al.* Measurement of the spin-forbidden dark excitons in $MoS_2$ and $MoSe_2$ monolayers. *Nature Communications* **2020**, *11*, 4037.

(29) Zhao, Y. C., Du, L. J., Yang, S. Q., Tian, J. P., Li, X. M., Shen, C., Tang, J., Chu, Y. B., Watanabe, K., Taniguchi, T. *et al.* Interlayer exciton complexes in bilayer $MoS_2$. *Physical Review B* **2022**, *105*, L041411.

(30) Lin, K.-Q., Faria Junior, P. E., Hübner, R., Ziegler, J. D., Bauer, J. M., Buchner, F., Florian, M., Hofmann, F., Watanabe, K., Taniguchi, T. *et al.* Ultraviolet interlayer excitons in bilayer $WSe_2$. *Nature Nanotechnology* **2023**, *19*, 196–201.

(31) Klein, J., Wierzbowski, J., Regler, A., Becker, J., Heimbach, F., Müller, K., Kaniber, M. & Finley, J. J. Stark Effect Spectroscopy of Mono- and Few-Layer $MoS_2$. *Nano Letters* **2016**, *16*, 1554-1559.

(32) Abraham, N., Watanabe, K., Taniguchi, T. & Majumdar, K. Anomalous Stark shift of excitonic complexes in monolayer $WS_2$. *Physical Review B* **2021**, *103*, 075430.

(33) Tagantsev, A. K., Stolichnov, I., Colla, E. L. & Setter, N. Polarization fatigue in ferroelectric films: Basic experimental findings, phenomenological scenarios, and microscopic features. *Journal of Applied Physics* **2001**, *90*, 1387-1402.

(34) Yang, S. M., Kim, T. H., Yoon, J. G. & Noh, T. W. Nanoscale Observation of Time-Dependent Domain Wall Pinning as the Origin of Polarization Fatigue. *Adv Funct Mater* **2012**, *22*, 2310-2317.

(35) Genenko, Y. A., Glaum, J., Hoffmann, M. J. & Albe, K. Mechanisms of aging and fatigue in ferroelectrics. *Mater Sci Eng B-Adv* **2015**, *192*, 52-82.

(36) Ievlev, A. V., Santosh, K. C., Vasudevan, R. K., Kim, Y., Lu, X. L., Alexe, M., Cooper, V. R., Kalinirr, S. V. & Ovchinnikova, O. S. Non-conventional mechanism of ferroelectric fatigue via cation migration. *Nature Communications* **2019**, *10*, 3064.

(37) Bian, R., He, R., Pan, E., Li, Z., Cao, G., Meng, P., Chen, J., Liu, Q., Zhong, Z., Li, W. *et al.* Developing fatigue-resistant ferroelectrics using interlayer sliding switching. *Science* **2024**, *385*, 57-62.

(38) Qin, B., Ma, C., Guo, Q., Li, X., Wei, W., Ma, C., Wang, Q., Liu, F., Zhao, M., Xue, G. *et al.* Interfacial epitaxy of multilayer rhombohedral transition-metal dichalcogenide single crystals. *Science* **2024**, *385*, 99-104.